\documentclass[12pt,preprint]{aastex}

\usepackage{natbib}
\usepackage[dvips]{color}

\usepackage{lscape}

\slugcomment{}

\shorttitle{Characterization of K2-19 Multi Transiting Planetary System}
\shortauthors{Narita et al.}

\begin{document}

\title{Characterization of the K2-19 Multiple-Transiting Planetary System
via High-Dispersion Spectroscopy, AO Imaging, and Transit Timing Variations}


\author{Norio Narita\altaffilmark{1,2,3}, Teruyuki Hirano\altaffilmark{4},
Akihiko Fukui\altaffilmark{5}, Yasunori Hori\altaffilmark{1,2},\\
Roberto Sanchis-Ojeda\altaffilmark{6,7}, Joshua N.\ Winn\altaffilmark{8},
Tsuguru Ryu\altaffilmark{2,3}, Nobuhiko Kusakabe\altaffilmark{1,2},\\
Tomoyuki Kudo\altaffilmark{9}, Masahiro Onitsuka\altaffilmark{2,3},
Laetitia Delrez\altaffilmark{10}, Michael Gillon\altaffilmark{10},\\
Emmanuel Jehin\altaffilmark{10}, James McCormac\altaffilmark{11},
Matthew Holman\altaffilmark{12}, Hideyuki Izumiura\altaffilmark{3,5},\\
Yoichi Takeda\altaffilmark{1}, Motohide Tamura\altaffilmark{1,2,13},
Kenshi Yanagisawa\altaffilmark{5}
}
\email{norio.narita@nao.ac.jp}


\altaffiltext{1}{Astrobiology Center, 2-21-1 Osawa, Mitaka, Tokyo, 181-8588, Japan}
\altaffiltext{2}{National Astronomical Observatory of Japan,
2-21-1 Osawa, Mitaka, Tokyo, 181-8588, Japan}
\altaffiltext{3}{SOKENDAI (The Graduate University for Advanced Studies),
2-21-1 Osawa, Mitaka, Tokyo, 181-8588, Japan}
\altaffiltext{4}{Department of Earth and Planetary Sciences, Tokyo Institute of
Technology, 2-12-1 Ookayama, Meguro-ku, Tokyo 152-8551, Japan}
\altaffiltext{5}{Okayama Astrophysical Observatory, National Astronomical
Observatory of Japan, Asakuchi, Okayama 719-0232, Japan}
\altaffiltext{6}{Department of Astronomy, University of California, Berkeley, CA 94720, USA}
\altaffiltext{7}{NASA Sagan Fellow}
\altaffiltext{8}{Department of Physics, and Kavli Institute for Astrophysics and Space Research,
Massachusetts Institute of Technology, Cambridge, MA 02139, USA}
\altaffiltext{9}{Subaru Telescope, 650 North A'ohoku Place, Hilo, HI 96720, USA}
\altaffiltext{10}{Institut d’Astrophysique et de G\'eophysique, Universit\'e de Li\`ege,
All\'ee du 6 Ao\^ut 17, Bat. B5C, 4000 Li\`ege, Belgium}
\altaffiltext{11}{Department of Physics, University of Warwick, Gibbet Hill Road, Coventry, CV4 7AL, UK}
\altaffiltext{12}{Smithsonian Astrophysical Observatory, 60 Garden St., Cambridge, MA 02138, USA}
\altaffiltext{13}{Department of Astronomy, The University of Tokyo, 7-3-1 Hongo, Bunkyo-ku, Tokyo,
113-0033, Japan}

\begin{abstract}
  K2-19 (EPIC201505350) is an interesting planetary system in which two
  transiting planets with radii $\sim7 R_{\oplus}$ (inner planet b)
  and $\sim4 R_{\oplus}$ (outer planet c) have orbits that are nearly
  in a 3:2 mean-motion resonance.  Here, we present results of
  ground-based follow-up observations for the K2-19 planetary system.
  We have performed high-dispersion spectroscopy and high-contrast
  adaptive-optics imaging of the host star with the HDS and HiCIAO on
  the Subaru 8.2m telescope.  We find that the host star is relatively
  old ($\geq8$ Gyr) late G-type star ($T_{\rm eff}\sim5350$ K, $M_{\rm
    s}\sim0.9$ $M_{\odot}$, and $R_{\rm s}\sim0.9$ $R_{\odot}$).  We
  do not find any contaminating faint objects near the host star which
  could be responsible for (or dilute) the transit signals.  We have
  also conducted transit follow-up photometry for the inner planet
  with KeplerCam on the FLWO 1.2m telescope, TRAPPISTCAM on the
  TRAPPIST 0.6m telescope, and MuSCAT on the OAO 1.88m telescope.  We
  confirm the presence of transit-timing variations, as previously
  reported by Armstrong and coworkers.  We model the observed
  transit-timing variations of the inner planet using the synodic
  chopping formulae given by Deck \& Agol (2015). We find two
  statistically indistinguishable solutions for which the period
  ratios ($P_{\rm c}/P_{\rm b}$) are located slightly above and below
  the exact 3:2 commensurability.  Despite the degeneracy, we derive
  the orbital period of the inner planet $P_{\rm b} \sim 7.921$ days
  and the mass of the outer planet $M_{\rm c} \sim 20 M_{\oplus}$.
  Additional transit photometry (especially for the outer planet) as
  well as precise radial-velocity measurements would be helpful to
  break the degeneracy and to determine the mass of the inner planet.
\end{abstract}


\section{Introduction}

Kepler's two-wheel mission, K2, has been in operation since 2014 \citep{2014PASP..126..398H}.
K2 observes a number of campaign fields in the ecliptic plane for about
83 days each, and releases photometric data at intervals of about 3
months.  Because many of the proposed target stars for K2 are somewhat
brighter than the planet-hosting stars discovered by the original
Kepler mission, the planets discovered by K2 are often good targets
for further characterization.  For this reason, planet searches based
on K2 photometric data and subsequent follow-up observations are being
conducted by many groups
\citep{2014PASP..126..948V,2015ApJ...809...25M,2015ApJ...806..215F,2015ApJ...804...10C,2015arXiv150404379S}.

EPIC201505350 (also designated K2-19) is one of the multi-transiting
planetary systems discovered in campaign field 1
\citep{2015ApJ...809...25M,2015ApJ...806..215F}.  An interesting
feature of this system is that two transiting planets (K2-19b and K2-19c) have
orbits that are close to a 3:2 mean-motion resonance (MMR).  The
formation of 3:2 MMR planet pairs through planetary migration has been
the subject of extensive theoretical investigations (e.g.,
\citealt{2005MNRAS.363..153P, 2008ApJ...687L.107R,
  2013ApJ...775...34O}); however, the number of 3:2 MMR pairs actually
discovered to date is still small.  Thus K2-19 offers the prospect of
a well-characterized example for theoretical studies of planet
formation, as one can determine the mass, radius, and density of both
planets, at least in principle.

The phenomenon of transit-timing variations (TTVs) can be used to
estimate the mass of planets that are in or near resonances
\citep{2005MNRAS.359..567A,2005Sci...307.1288H}.  Recent theoretical
studies have shown that the TTV of near-MMR planet pairs can be
calculated analytically
\citep{2012ApJ...761..122L,2014ApJ...790...58N,2015ApJ...802..116D},
facilitating the analysis.  As both K2-19b and c are transiting, it
may be possible to determine or constrain the mass of both planets
through TTV monitoring.  Ground-based transit follow-up observations
for this system are very important, since the monitoring period of K2
was limited to about 83 days.  Even though no TTV were detected during
the interval of the K2 observations, recently
\citet{2015A&A...582A..33A} reported on subsequent observations of
K2-19b using the NITES 0.4m telescope and showed that the planet
experienced a large TTV.  They used a formula by
\citet{2012ApJ...761..122L} to put an upper bound of $\sim$300
M$_{\oplus}$ on the masses of K2-19b and c.  Further transit follow-up
observations are needed to place a more stringent upper bound, or to
determine the masses of the planets.

Another interesting feature of K2-19 is that the inner planet is a
super-Neptune with the radius of about 7~$R_{\oplus}$ (or 0.7~$R_{\rm Jup}$).
Based on an examination of the NASA Exoplanet Archive \citep{2013PASP..125..989A},
planets in that size range appear to be relatively infrequent among
the sample of known transiting planets. Only a small number of
super-Neptune-sized planets, such as HATS-7b ($V$=13.34,
\citealt{2015arXiv150701024B}) and HATS-8b ($V$=14.03,
\citealt{2015AJ....150...49B}), have been discovered.  Further, K2-19
is a unique MMR system in that the inner planet (super-Neptune-size)
is about 1.7 times bigger than the outer planet (Neptune-size).  This
is not the case for the majority of 3:2 MMR planet pairs discovered by
the Kepler \citep{2013ApJ...763...41C}, as discussed later in this paper.
Those facts make it interesting to investigate the origin of K2-19 system as well
as the internal and atmospheric compositions of both planets.

Another issue regarding this system is that the stellar parameters of
the host star, such as the stellar mass and radius, have been characterized
with only low signal-to-noise ratio (SNR of $\sim 25$),
moderately high spectral resolution ($R \sim 39,000$) spectra
(see \citealt{2015A&A...582A..33A}).
Thus there is room for improvement in the estimation of
stellar parameters with higher SNR and higher spectral resolution spectra.
Furthermore, the possibility has not yet been excluded
that there is a faint neighboring star with the photometric aperture
of the K2 detector, which could be responsible for the transit signals
or which could affect the observed amplitude of the signals.  Both
characterization of the host star with high dispersion spectroscopy
and high-contrast direct imaging are important for a precise determination
of the masses and radii of the planets.

Motivated by the preceding considerations, we have conducted three
types of ground-based follow-up observations.  The first one is
high-dispersion spectroscopy to characterize the host star.  The
second one is high-contrast adaptive-optics (AO) imaging to check on
any possible contamination from a faint companion star.  The third one
is ground-based time-series photometry of transits, to measure
additional transit times and enhance the TTV analysis.  The rest of
this paper is organized as follows.  We introduce our observations and
reduction procedures in section~2.  We describe our analysis methods
and results in section~3.  Based on the derived transit parameters, we
further analyze the TTVs of the planets in section~4.1.  We also
discuss possible internal compositions and origins of the planets from
a theoretical point of view in section~4.2 and 4.3.  Finally, we
summarize our study in section~5.

\section{Observations and Reductions}

\subsection{Subaru 8.2m Telescope / HDS}

In order to improve on estimates of the stellar parameters, we
observed K2-19 with High Dispersion Spectrograph (HDS) on the Subaru
8.2m telescope on 2015 May 29 (UT).  To maximize the SNR
as well as to achieve a sufficient spectral resolution, we
employed the Image Slicer \#2 \citep[][; $R\sim
80,000$]{2012PASJ...64...77T} and the standard I2a setup,
simultaneously covering the spectral range between 4950-7550
$\mathrm{\AA}$ with the two CCD chips.  The raw spectrum was subjected
to the standard IRAF\footnote{The Image Reduction and Analysis
  Facility (IRAF) is distributed by the National Optical Astronomy
  Observatory, which is operated by the Association of Universities
  for Research in Astronomy (AURA) under a cooperative agreement with
  the National Science Foundation.}  procedures to extract the
one-dimensional (1D) spectrum.  The wavelength scale was established
by observations of the comparison Thorium-Argon lamp taken during
evening and morning twilight.  The exposure time was 20 minutes.
The 1D spectrum has a SNR of $\sim 70$ per pixel in the vicinity of the sodium D lines.

\subsection{Subaru 8.2m Telescope / HICIAO \& AO188}

We observed K2-19 in the $H$ band with the HiCIAO
\citep{2006SPIE.6269E..28T} combined with the AO188 (188 element
curvature sensor adaptive optics system:
\citealt{2008SPIE.7015E..25H}), mounted on the 8.2m Subaru Telescape
on 2015 May 8 (UT).  We used the target star itself as a natural guide
star for AO188, and employed an atmospheric dispersion corrector (ADC)
to prevent the star from drifting on the detector due to differential
refraction between the visible and near-infrared bands.  The field of
view (FOV) was $20'' \times 20''$, and the typical AO-corrected seeing
was $\sim0\farcs1$ on the night of our observations.  The observations
were conducted in the pupil tracking mode to enable the angular
differential imaging (ADI; \citealt{2006ApJ...641..556M}) technique.
We took 35 object frames with an exposure time of 30~s.  The total
exposure time was 17.5~min.  As a first step in the reduction, we removed a
characteristic stripe bias pattern from each image
\citep{2010SPIE.7735E..30S}. Then, bad pixel and flat field correction
were performed.  Finally, the image distortions were corrected, using
calibration images of the globular cluster M5 that were obtained on
the same night.

\subsection{FLWO 1.2m Telescope / KeplerCam}

We observed one transit of K2-19 on 2015 January 28 (UT) with the 1.2m
telescope at the Fred Lawrence Whipple Observatory (FLWO) on Mt.\
Hopkins, Arizona.  We used Keplercam, which is equipped with a $4096
\times 4096$ pixel CCD with a $23\farcm1$$\times$$23\farcm1$ FOV.  We
observed through a Sloan $i'$ filter.
The exposure time was 30 s. Debiasing and flat-fielding
(using dome flats) were performed using standard IRAF procedures.
Aperture photometry was performed with custom routines written in the
Interactive Data Language (IDL).  We selected the final aperture size
of 7 pixels in the 2$\times$2 binning configuration, which means the
radius of the aperture is about $5"$.
The sky level per pixel was estimated from the median value in
an annulus surrounding the aperture, with a radius that is
about twice the aperture radius. 
The time of each exposure was recorded in UT, and the systematic error of
the recorded time with respect to the standard clock was
much smaller than the statistical uncertainty for the mid-transit time.

\subsection{TRAPPIST 0.6m Telescope / TRAPPISTCAM}

One transit of K2-19b was observed on the night of 2015 Feb 28 (UT)
with the 0.6m TRAPPIST robotic telescope (TRAnsiting Planets and
PlanetesImals Small Telescope), located at ESO La Silla Observatory
(Chile).  TRAPPIST is equipped with a thermoelectrically-cooled
2K$\times$2K CCD, which has a pixel scale of $0\farcs65$ that
translates into a $22' \times 22'$ FOV.  For details of TRAPPIST, see
\citet{2011EPJWC..1106002G} and \citet{2011Msngr.145....2J}.  The
transit was observed in an Astrodon Exoplanet (blue-blocking) filter
that has a transmittance over 90\% from 500~nm to beyond 1000~nm.
The exposure time was 10 s. The time of each exposure was recorded
in HJD$_{\rm UTC}$.
During the run, the positions of the stars on the chip were maintained
to within a few pixels thanks to a software guiding system that
regularly derives an astrometric solution for the most recently
acquired image and sends pointing corrections to the mount if needed.
After a standard pre-reduction (bias, dark, and flat-field
correction), the stellar fluxes were extracted from the images using
the IRAF / DAOPHOT aperture photometry software
\citep{1987PASP...99..191S}.  After testing several sets of reduction
parameters, we chose the one giving the most precise photometry for
the stars of similar brightness as the target.  Differential
photometry of the target star was performed relative to a
selected set of reference stars.
The set of reference stars was chosen as the one that
gives the lowest root-mean-square (rms) transit light curve.
It consists of nine stable stars of similar brightness and color to the target.

\subsection{OAO 1.88m Telescope / MuSCAT}

We observed one transit of K2-19b with a new multi-color camera, named
MuSCAT, installed on the 1.88m telescope in Okayama Astrophysical
Observatory (OAO) on 2015 April 25 (UT).  On that night, a full
transit was predicted to be observable based on the ephemeris of
\citet{2015ApJ...806..215F}.  However, based on the updated ephemeris
of \citet{2015A&A...582A..33A} (taking TTVs into account) only a
partial transit was predicted to be observable.  MuSCAT equips 3 CCD
cameras and has a capability of 3-color simultaneous imaging in
$g'_2$, $r'_2$, and $z_{s,2}$ bands of Astrodon Sloan Gen 2 filters
\citep{doi:10.1117/1.JATIS.1.4.045001, 2015arXiv151003997F}.
Each CCD camera has a
$6\farcm1$$\times$$6\farcm1$ FOV, with a pixel scale of about
$0\farcs358$.  For the $g'_2$ and $r'_2$ band observations, we
employed the high-speed readout mode (2 MHz, corresponding readout
time of about 0.58 s and readout noise of $\sim10$ e$^-$).  For the
$z_{s,2}$ band observations we used the low-speed readout mode (100
kHz, corresponding to readout time of about 10 s and readout noise of
$\sim4$ e$^-$) because at that time the $z_{s,2}$-band CCD was
affected by excess readout noise (over $20$ e$^-$) in the
high-speed readout mode\footnote{The CCD has since been repaired and
  the readout noise problem has been fixed.}.  The exposure times
were 60 s, 30 s, and 60 s for $g'_2$, $r'_2$, and $z_{s,2}$ bands,
respectively.  Bias subtraction, flat-fielding, and aperture
photometry are performed by a customized pipeline
\citep{2011PASJ...63..287F}.  Aperture radii are selected as 17, 18,
and 19 pixels for $g'_2$, $r'_2$, and $z_{s,2}$ bands, respectively.
Sky level is measured in the annulus with the inner radius of 45
pixels and the outer radius of 55 pixels.  For differential
photometry, we select a reference star UCAC4 453-052399 ($g$=12.53,
$r$=12.11, $i$=11.96, $J$=11.08), which is slightly brighter than the
target star ($g$=13.36, $r$=12.76, $i$=12.57, $J$=11.60).  The
aperture radii and the reference star were chosen so that
rms of residuals for trial transit fittings are
minimized.

\subsection{NITES 0.4m telescope}

For a more comprehensive transit analysis, we also analyze the
previously published data from the NITES (Near Infra-red Transiting
ExoplanetS: \citealt{2014MNRAS.438.3383M}) 0.4m telescope in the same
manner as our own data. The NITES data are identical with those
presented by \citet{2015A&A...582A..33A}.  The transit observed by
NITES was the same as the one observed by TRAPPIST.

\section{Analyses and Results}

\subsection{Host Star Parameters from High Dispersion Spectroscopy}
\label{HDSresult}

Using the high-resolution spectrum obtained with Subaru/HDS, we
extract the atmospheric parameters (the effective temperature
$T_\mathrm{eff}$, surface gravity $\log g$, metallicity [Fe/H], and
microturbulent velocity $\xi$) by measuring the equivalent widths of
iron I/II lines \citep{2002PASJ...54..451T}, which are widely
distributed throughout the observed wavelength region.  We then
translate those atmospheric parameters into estimates of the stellar
mass and radius, using the Yonsei-Yale stellar-evolutionary models
\citep{2001ApJS..136..417Y}.  Finally, the projected stellar rotation
velocity $V\sin I_s$ is estimated by fitting the observed spectrum
with the theoretically-generated intrinsic stellar absorption line
profile, convolved with a kernel taking into account rotational and
macroturbulent broadening \citep{2005oasp.book.....G} and the
instrumental profile (IP) of Subaru/HDS.
The theoretical synthetic spectrum was taken from ATLAS9 model
(a plane-parallel stellar atmosphere model in LTE: \citealt{1993KurCD..13.....K}).
The result of the spectroscopic analysis is summarized in Table~\ref{table:HDS}.
See \citet{2014ApJ...783....9H} for details on how the uncertainties in
these spectroscopic parameters are determined.  The derived parameters
are consistent with those reported by \citet{2015A&A...582A..33A}, and
have a higher precision by about an order of magnitude thanks to
higher SNR spectra.  We also find that the host star is likely older
than $\sim$8 Gyr, which is far older than the typical planet formation
and migration timescale ($\sim10^7$ yr, see also Section 4.2).

\subsection{Constraint on Contaminating Stars from AO Imaging}

\begin{figure*}[ht]
		\begin{center}
			\includegraphics[width=8.5cm]{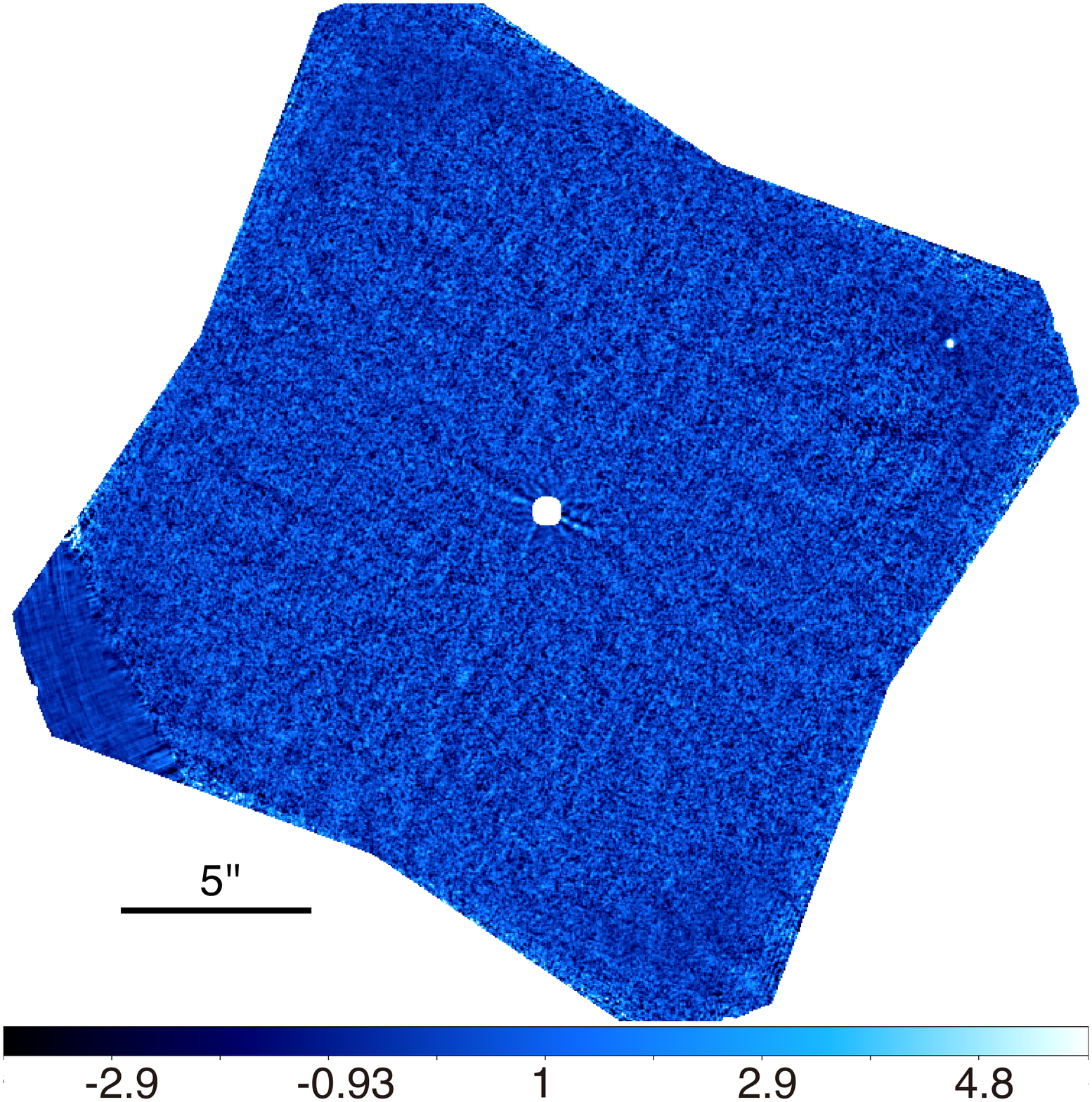} 
			\caption{A signal-to-noise ratio (SNR) map around K2-19 processed by the LOCI algorithm.
			North is up and East is left.
			\label{snmap}}
			\includegraphics[width=8.5cm]{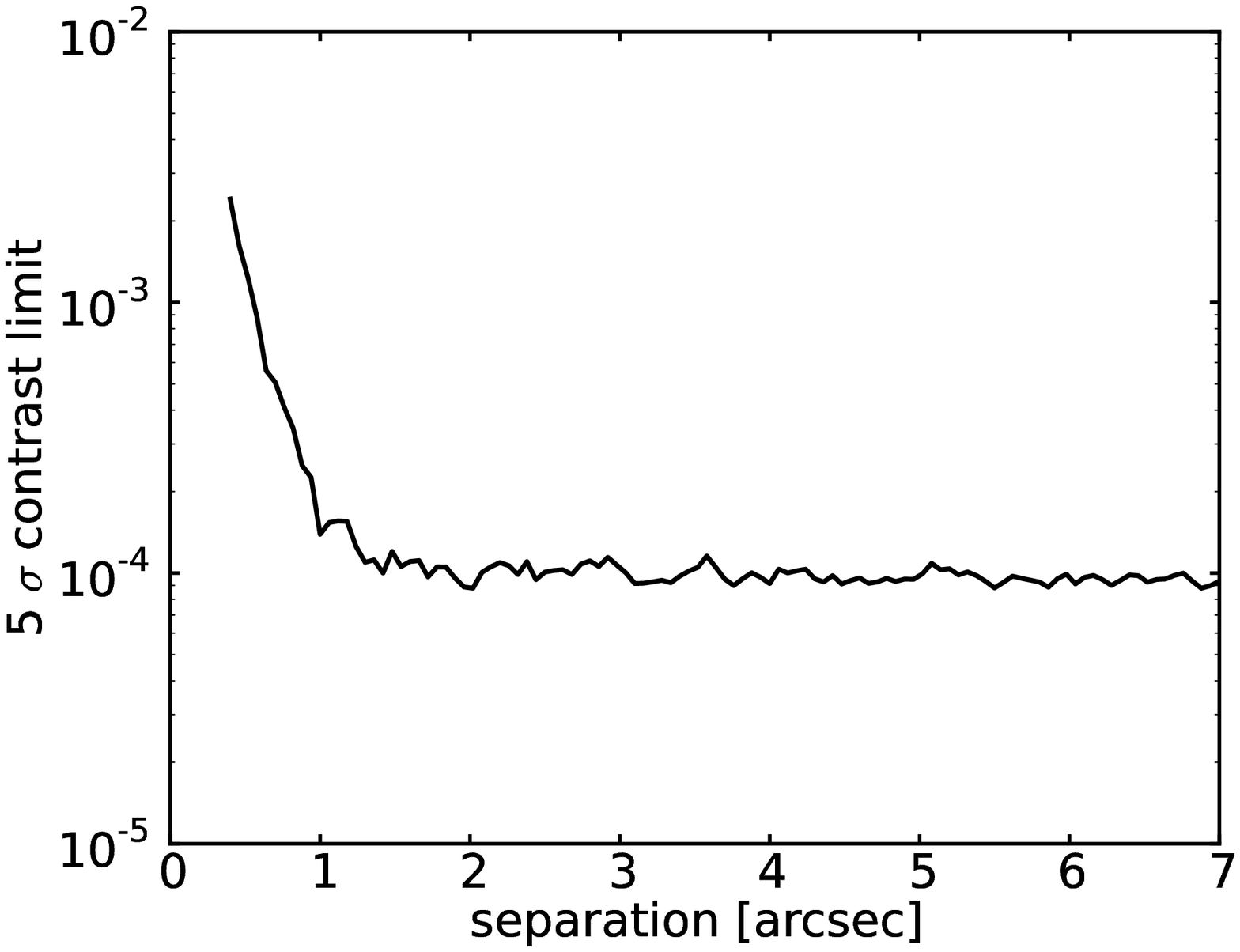} 
			\vspace{35mm}
			\caption{A 5$\sigma$ contrast limit curve around K2-19.
			\label{contrast}}
		\end{center}
	\end{figure*}

We employ the locally optimized combination of images (LOCI: \citealt{2007ApJ...660..770L})
algorithm to maximize the efficiency of the ADI technique and to search for
fainter objects in the inner region around K2-19.
We estimate our achieved contrast ratio as follows.
The final image of a LOCI pipeline is convolved with the PSF of the unsaturated image.
The 1$\sigma$ limit on the contrast ratio
is defined as the ratio of the standard deviation of the pixel values
inside an annulus of width 6 pixels to the stellar flux in the unsaturated image.
Then the contrast ratio is corrected for a self-subtraction effect, which is estimated from
the recovery rate of injected signals.
Figure~\ref{snmap} shows a signal-to-noise ratio (SNR) map of a LOCI-reduced image
around K2-19 (North is up and East is to the left).
There are no other faint sources around K2-19 with SNR exceeding 5, except for
a faint source located at the west-northwest edge of the image ($\sim$11$\farcs$5 from the center).
The faint source has a contrast of $\sim4\times10^{-4}$ relative to the brightness of K2-19
and locates outside the Kepler's PSF. Thus the presence of the faint source does not affect
the light curves from K2 and other ground-based telescopes.
We plot the 5$\sigma$ contrast limit in Figure~\ref{contrast}.
Contrast limits of $\sim2\times10^{-3}$ and $\sim$10$^{-4}$ are achieved
at distances of at 0$\farcs$4 and 1'' from the host star, respectively.
Thus we have not identified any contaminating faint object which can dilute or mimic
the transits of K2-19b and K2-19c. 

\subsection{Reanalysis of K2 Light Curve}

\begin{figure*}[ht]
		\begin{center}
			\includegraphics[width=8.5cm]{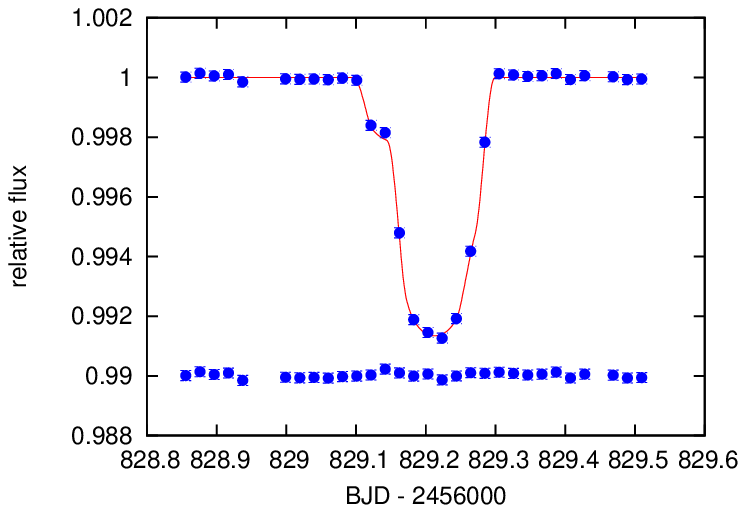}
			\includegraphics[width=8.5cm]{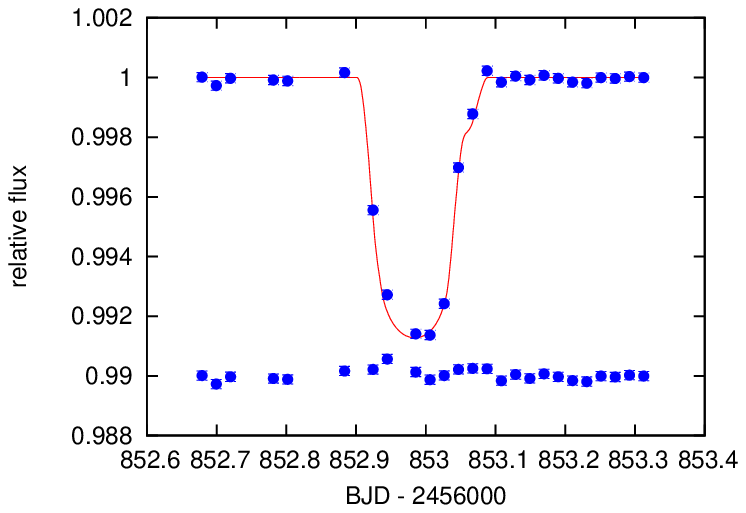}
			\includegraphics[width=8.5cm]{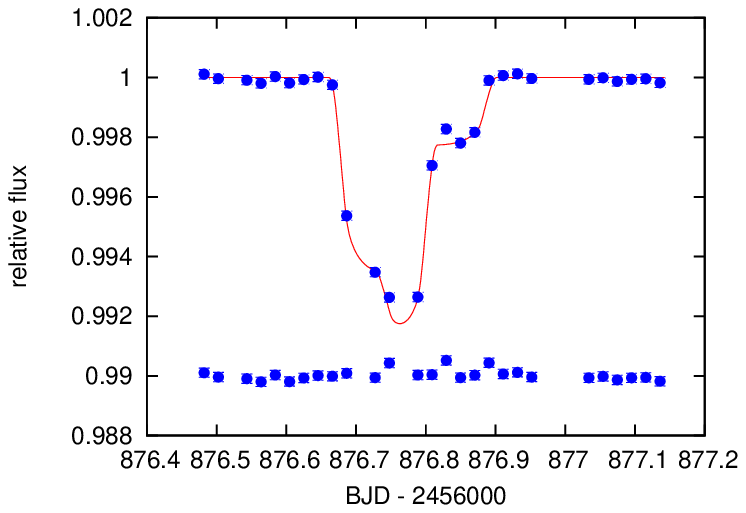} 
			\vspace{12mm}
			\caption{Mutual transit events observed in K2 data.
			Out-of-transit regions are normalized by the ESPRINT pipeline \citep{2015arXiv150404379S}.
			Blue points show the light curve data and the red solid lines represent the best-fit models.
			Residuals are plotted with vertical offset by -0.01.
			}
			\label{K2mutual}
		\end{center}
	\end{figure*}

We analyze the light curve for K2-19 that has been produced from the raw pixel data
by the ESPRINT collaboration. See \citet{2015arXiv150404379S} for a
description of the procedures for extracting the light curve from the K2 pixel data.
We fit the K2 light curve by the procedure below and derive the best-fit system parameters
along with mid-transit times.
First, we separate the entire light curve into fourteen segments, each
of which spans a transit.
Three of the segments involve double transits (see Figure~\ref{K2mutual}),
during which two planets simultaneously transit the host star.
We note that \citet{2015A&A...582A..33A} did not model those mutual transit events.
Each segment includes the data within $\sim 4.5$ hours of the predicted ingress/egress times
(based on the ephemeris by \citealt{2015A&A...582A..33A}), which is
sufficient to span the transit event as well as some time beforehand and afterward.
For each segment, we compute the standard deviation of the out-of-transit fluxes,
and adopt the standard deviation as an initial estimate of the
uncertainty in each flux value.

All the light curve segments are fitted simultaneously with the common system
parameters along with mid-transit time(s) for each segment.
We compute the posterior distributions for those parameters based on
the Markov Chain Monte Carlo (MCMC) algorithm,
assuming that the likelihood is proportional to $\exp(-\chi^2/2)$ where
\begin{eqnarray}
\label{eq:chisq}
\chi^2 = \sum_{i}\frac{(f_\mathrm{LC,obs}^{(i)}-f_\mathrm{LC, model}^{(i)})^2}{\sigma_\mathrm{LC}^{(i)2}},
\end{eqnarray}
where $f_\mathrm{LC,obs}^{(i)}$ and $\sigma_\mathrm{LC}^{(i)}$ are
the $i$-th observed K2 flux and its error, respectively. 
We employ the transit model by \citet{2009ApJ...690....1O} integrated over
the 29.4-minute averaging time of the {\it Kepler} data.
We neglect the impact of any possible planet-planet eclipses \citep{2012ApJ...759L..36H}
on the light curve shape, due to the rather sparse sampling of the K2 data.
Finally, to obtain the model flux $f_\mathrm{LC, model}^{(i)}$,
the integrated light curve model for each segment is multiplied
by a second-order polynomial function of time, representing the
longer-timescale flux variations at the time of the transit.
Therefore, the adjustable parameters in our model are (for each
planet) the scaled semi-major axis $a/R_s$,
the transit impact parameter $b$, and the planet-to-star radius ratio $R_p/R_s$;
the quadratic limb-darkening parameters $u_1$ and $u_2$;
the mid-transit time(s) $T_c$ for each segment; and the coefficients of the second-order polynomials.
Because the sparse sampling of the K2 data relative to the timescale of ingress and egress
prohibits us from constraining the orbital eccentricity $e$ of either planet
based only on the transit light curve \citep{2012ApJ...756..122D},
we simply assume $e=0$ for both planets. 

Following the procedure described in \citet{2015ApJ...799....9H},
we compute the posterior distributions for the parameters listed above;
we first optimize Equation (\ref{eq:chisq}) using Powell's conjugate direction method,
allowing all the relevant parameters to vary.
At this point, we compute the level of time-correlated noise $\beta$
(so-called red noise: \citealt{2006MNRAS.373..231P,2008ApJ...683.1076W}) for each segment,
and inflate the errors in each segment by $\beta$.
We then fix the coefficients of the polynomial functions to the optimized values
and perform the MCMC computation starting from the initial best-fit values for the other parameters.
The step size for each parameter is iteratively optimized such that the final acceptance
ratio over the whole chain is between 15--35\%. We extend the chains
to $10^7$ links, given the relatively large number of free parameters.
We employ the median, and 15.87 and 84.13 percentiles of the marginalized posterior
distribution of each parameter to convey the representative value and its $\pm 1\sigma$ errors.
As a check on convergence, we repeat the analysis after
changing the initial input values for the system parameters,
and find no dependence on the initial values.
We report the values and uncertainties of the basic transit parameters in Table~\ref{table:K2},
and we report the mid-transit times as well as the $\beta$ factor for each segment in Table~\ref{table:K2Tc}.
Combining the $R_{\rm p}/R_{\rm s}$ values with the $R_{\rm s}$ value listed in Table~\ref{table:HDS},
we estimate the planetary radii of K2-19b and K2-19c as $R_{\rm p,b} = 7.34 \pm 0.27 R_{\oplus}$
and $R_{\rm p,c} = 4.37 \pm 0.22 R_\oplus$, respectively.
The planetary radii are also presented in Table~\ref{table:K2}.

\subsection{Modeling of New Transit Light Curves}

Our procedure for modeling the ground-based transit light curves follows
\citet{2013ApJ...773..144N}.
First, the time stamps of the photometric data are placed
onto the $\mathrm{BJD_{TDB}}$ system using the code by \citet{2010PASP..122..935E}.
We check all the light curves by eye and eliminate obvious outliers.
We then fit the transit light curves simultaneously, using
an analytic transit light curve model and various choices to describe
the more gradual out-of-transit (baseline) flux variations.

We adopt baseline model functions $F_{\rm oot}$ described as follows,
\begin{eqnarray*}
F_{\rm oot} &=& k_0 \times 10^{-0.4\Delta m_{\rm cor}},\\
\Delta m_\mathrm{cor} &=& \sum k_i X_i,
\end{eqnarray*}
where $k_0$ is the normalization factor,
$\{{\bf X}\}$ are observed variables, and $\{{\bf k}\}$ are the coefficients \citep{2013ApJ...770...95F}.
In order to select the most appropriate baseline models for the light curves,
we adopt the Bayesian Information Criteria (BIC: Schwarz 1978).
The BIC value is given by $\mathrm{BIC} \equiv \chi^2 + k \ln N$,
where $k$ is the number of free parameters, and $N$ is the number of data points. 
For the variables $\{{\bf X}\}$, we test various combinations of time ($t$), airmass ($z$),
the relative centroid positions in x ($dx$) and y ($dy$), and the sky background counts ($s$).
In addition, we also allowed an additional factor $k_{\rm flip}$ to account for
a change of the normalization factor after the meridian flip for the TRAPPIST light curve.
Based on minimum BIC, we adopt $k_0$. $t$ and $z$ to be fitting parameters for FLWO and MuSCAT;
$k_0$, $z$, and $k_{\rm flip}$ for TRAPPIST; and $k_0$ and $t$ for NITES, respectively.

For the transit light curve model, we employ a customized code \citep{2007PASJ...59..763N}
that use the analytic formula by \citet{2009ApJ...690....1O}.
The formula is equivalent to that of \citet{2002ApJ...580L.171M} when using the quadratic limb-darkening law,
$I(\mu) = 1 - u_1 (1-\mu) - u_2 (1-\mu)^2$,
where $I$ is the intensity and $\mu$ is the cosine of the angle between
the line of sight and the line from the position of the stellar surface to the stellar center.
We refer the tables of quadratic limb-darkening parameters by \citet{2013A&A...552A..16C},
and compute allowed $u_1 + u_2$ and $u_1 - u_2$ values
for $T_{\rm eff} =$ 5300 K or 5400 K and $\log g =$ 4.0 or 4.5,
based on the stellar parameters presented in Table~\ref{table:HDS}.
We adopt uniform priors for $u_1 + u_2$ and $u_1 - u_2$ as follows:
[0.59, 0.61] and [0.24, 0.34] for FLWO $i'$,
[0.77, 0.85] and [0.67, 0.82] for MuSCAT $g'_2$,
[0.68, 0.70] and [0.35, 0.46] for MuSCAT $r'_2$,
[0.52, 0.55] and [0.17, 0.26] for MuSCAT $z_{s,2}$, and
[0.67, 0.70] and [0.38, 0.50] for TRAPPIST and NITES.
We assume an orbital period $P = 7.921$ days and
a reference epoch for the transits $T_{\rm c,0} = 2457082.6895$,
the values reported by \citet{2015A&A...582A..33A} based on the NITES data.
We note that these assumptions have little impact on the resultant transit parameters,
because we allow $T_{\rm c}$ to be a free parameter, and the uncertainty in $P$ is negligible. 
To constrain the mid-transit times precisely even without complete
coverage of the entire transit event,
we place an {\it a priori} constraint on the total transit duration, $T_{14} = 0.1365 \pm 0.0017$,
based on our K2 analysis.
We allow $R_{\rm p}/R_{\rm s}$, $a/R_{\rm s}$, and the orbital
inclination $i$ to be free parameters.

\begin{figure*}[ht]
		\begin{center}
			\includegraphics[width=12cm]{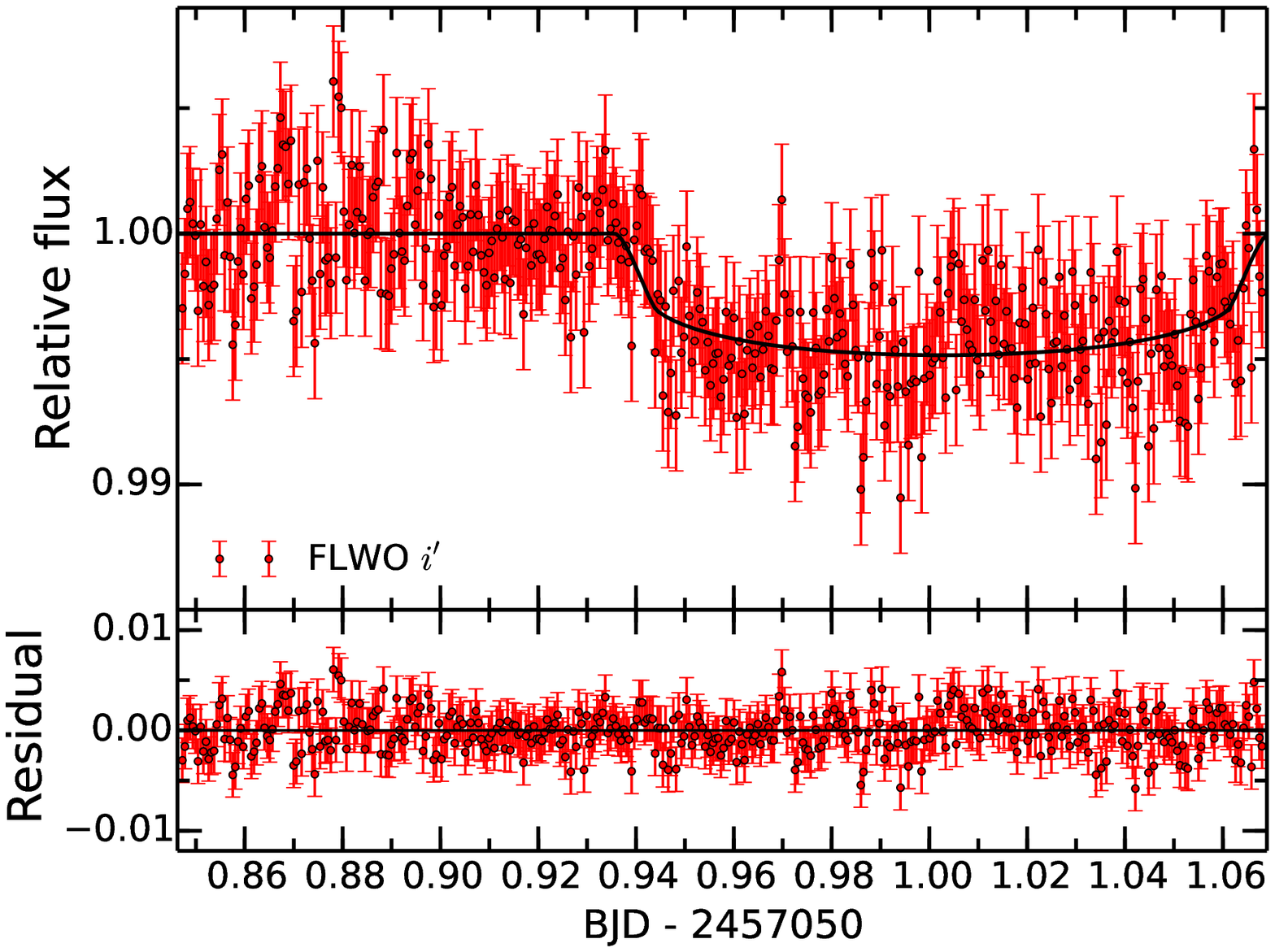} 
			%\vspace{40mm}
			\caption{A transit light curve obtained with FLWO 1.2m telescope.}
			\label{lcFLWO}
			%\vspace{-35mm}
			\includegraphics[width=12cm]{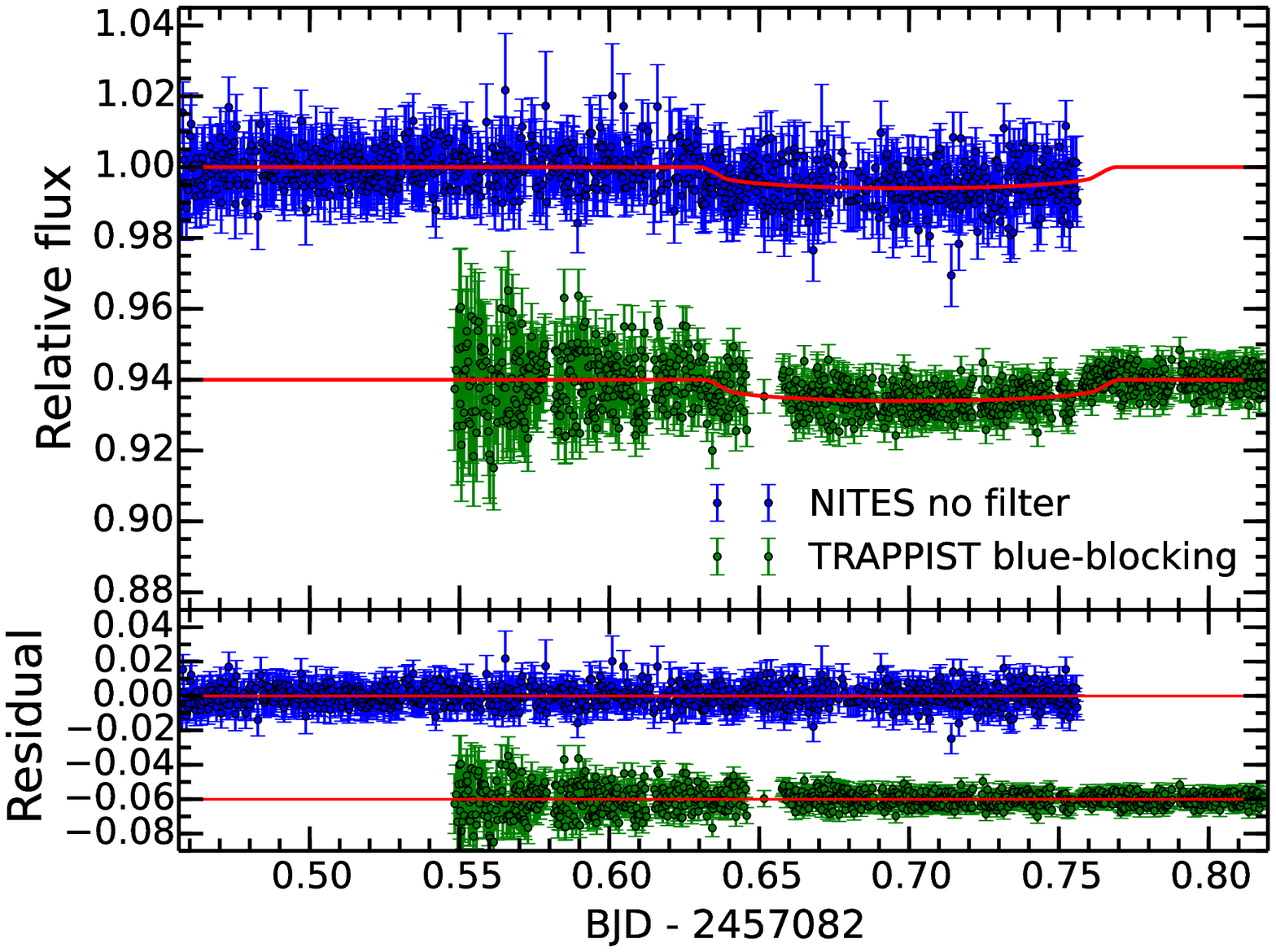} 
			%\vspace{40mm}
			\caption{Transit light curves taken by TRAPPIST 0.6m and NITES 0.4m telescopes.}
			\label{lcTRNI}
		\end{center}
	\end{figure*}

\begin{figure*}[ht]
		\begin{center}
			\includegraphics[width=12cm]{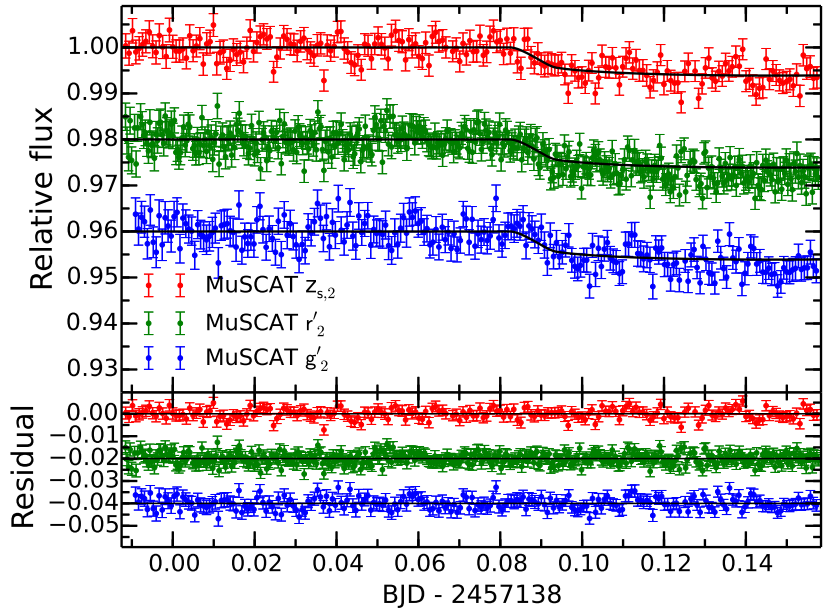} 
			%\vspace{40mm}
			\caption{Same as Fig.~\ref{lcTRNI}, but by MuSCAT on OAO 1.88m telescope.}
			\label{lcMuSCAT}
		\end{center}
	\end{figure*}

First we optimize free parameters for each light curve, using the AMOEBA algorithm
\citep{1992nrca.book.....P}.
The penalty function is also given by Eq.~(\ref{eq:chisq}),
where $f_{\rm LC,model}^{(i)}$ is a combination of
the baseline model and the analytic transit formula mentioned above.
Then if the reduced $\chi^2$ is larger than unity,
we rescale the photometric errors of the data
such that the reduced $\chi^2$ for each light curve becomes unity.
We also estimate the level of time-correlated noise for each light curve,
and further rescale the errors by multiplying $\beta$ factors listed in Table~\ref{table:ground}.
Finally, we use the MCMC method \citep{2013ApJ...773..144N}
to evaluate values and uncertainties of the free parameters.
For the FLWO, TRAPPIST, and NITES light curves,
we fit each light curve independently.
We create 5 chains of 10$^6$ points for each light curve,
and discard the first 10$^5$ points from each chain (the ``burn-in'' phase).
The jump sizes are adjusted such that the acceptance ratios are 20--30\%.
For the OAO/MuSCAT light curves, we fit the 3-band data
simultaneously,
requiring consistency in the parameters $T_{\rm c}$, $i$, and $a/R_{\rm s}$.
We create 5 chains of 3$\times$10$^6$ points for MuSCAT light curves,
and discard the first 3$\times$10$^5$ points from each chain.
The acceptance ratios are set to about 25\%.
Table~\ref{table:ground} lists the median values and $\pm1\sigma$ uncertainties, which are defined
by the 15.87 and 84.13 percentile levels of the merged posterior distributions.
The baseline corrected transit light curves are plotted in Figure~\ref{lcFLWO} (FLWO),
Figure~\ref{lcTRNI} (TRAPPIST and NITES), and Figure~\ref{lcMuSCAT} (OAO).

\section{Discussion}

\subsection{Transit Timing Variations and the Mass of the Outer Planet}

Since K2-19b and K2-19c are close to the 3:2 MMR, it is not surprising
that large TTVs have been observed in this system.
To explain the observed mid-transit times, we adopt the analytic formulae of TTV by
\citet{2015ApJ...802..116D} and \citet{2014ApJ...790...58N}, who considered the synodic chopping effect.
Successful modeling of the synodic chopping effect enables us to estimate the mass of a perturbing body.
In particular the resulting mass determination is not subject to
the mass-eccentricity degeneracy that was discussed by \citet{2012ApJ...761..122L},
who derived analytic formulae for a pair of planets near a first-order MMR.
Although the chopping analytic formulae are only applicable to systems
with nearly circular and coplanar orbits,
those requirements are likely fulfilled in this case,
given the presence of mutual transits and the requirement for
long-term dynamical stability.

The chopping formulae predict the TTVs of the inner and outer planets as follows:
\begin{eqnarray}
\delta t_b &=& \sum_{j=1}^\infty {P_b \over 2\pi} \mu_c f_b^{(j)}(\alpha) \sin{ \psi_j},\\
\delta t_c &=& \sum_{j=1}^\infty {P_c \over 2\pi} \mu_b f_c^{(j)}(\alpha) \sin{\psi_j},
\end{eqnarray}
where $\mu_{b,c}$ are the planet-to-star mass ratio ($M_{b,c}/M_s$) respectively, and
\begin{eqnarray}
f_b^{(j)}(\alpha) &=& -\alpha \frac{j(\beta^2+3) b_{1/2}^{j}(\alpha)+2 \alpha \beta \frac{\partial}{\partial \alpha} b_{1/2}^{j}(\alpha)-\alpha \delta_{j,1} (\beta^2+2\beta+3) }{\beta^2 (\beta^2-1)},\\
f_c^{(j)}(\alpha) &=&  \frac{j(\kappa^2+3) b_{1/2}^{j}(\alpha)+2 \kappa (\alpha \frac{\partial}{\partial \alpha} b_{1/2}^{j}(\alpha)+b_{1/2}^{j}(\alpha))-\alpha^{-2} \delta_{j,1} (\kappa^2-2\kappa+3) }{\kappa^2 (\kappa^2-1)},\\
b_{1/2}^j(\alpha) &=& \frac{1}{\pi} \int_0^{2\pi} \frac{\cos{(j \theta)}}{\sqrt{1-2\alpha \cos{\theta}+\alpha^2}}d \theta.
\end{eqnarray}
Note that $P_{b,c}$ and $\mu_{b,c}$ are defined as non-negative variables.
The parameters $\alpha$, $\beta$, and $\kappa$ are defined as,
$\alpha = a_b/a_c$, $\beta = j(n_b-n_c)/n_b$, and $\kappa = j(n_b-n_c)/n_c$,
where $a_{b,c}$ and $n_{b,c}$ are the semi-major axes and the mean motions of the planets.
Those parameters can be expressed by the periods of the planets:
$\alpha \simeq (P_b/P_c)^{2/3}$ (using Kepler's third law and
neglecting the planet masses in comparison to the stellar mass),
$\beta = j (1- P_b/P_c)$, and $\kappa = j (P_c/P_b - 1)$.
Note that $\delta_{j,1}$ represents Kronecker's delta, which is 1 when
$j=1$ and 0 for $j>1$.
When the time of conjunction ($t_{\rm conj}$) is used as the origin of
the time scale,
the phase $\psi_j$ can be expressed simply as
\begin{eqnarray}
\psi_j = 2\pi t j (1/P_b -1/P_c) \equiv 2 \pi t j / P_{\rm syn},
\end{eqnarray}
where $P_{\rm syn}$ is the synodic period.
For the current system, we derive $t_{\rm conj} = 2456852.9344 \pm 0.0022$ BJD$_{\rm TDB}$
from one of the mutual transit events,
assuming that both planets are orbiting in the same direction.
Consequently, the observed mid-transit times ($T_c$) for the planets can be modeled by
6 free parameters $T_c (0)_{b,c}$, $P_{b,c}$, and $\mu_{b,c}$ as:
\begin{eqnarray}
T_c (E_b)_b &=& T_c (0)_b + P_b \times E_b + \delta t_b,\\
T_c (E_c)_c &=& T_c (0)_c + P_c \times E_c + \delta t_c,
\end{eqnarray}
where $E_{b,c}$ are the transit epochs of the planets with the origins at the first K2 transits.

We first try to find an optimal parameter set for the 6 parameters
($T_c (0)_{b}$, $T_c (0)_{c}$, $P_{b}$, $P_{c}$, $\mu_{b}$ and $\mu_{c}$) by minimizing
\begin{eqnarray}
\chi^2 = \sum_{i}\frac{(T_{\rm c,obs}^{(i)}-T_{\rm c,model}^{(i)})^2}{\sigma_{T_{\rm c}}^{(i)2}},
\end{eqnarray}
using the AMOEBA algorithm.  In evaluating Equations~(2) and (3), we
truncate the series at $j=7$ to reduce the computational cost.  We
model the observed the mid-transit times for both planets listed in
Table~\ref{table:K2Tc} and \ref{table:ground} (21 in total) using the
equations above.  However, we find that the AMOEBA algorithm does not
converge.  Note that we have tested the effects of including higher orders of $j$ in
the calculation, but the result remains unchanged.  Apparently there are
multiple local minima of $\chi^2$ in the parameter space, which are
almost equally favored.  This result is quite understandable: we have
not observed any additional transits of planet c since the end of the
K2 campaign, and hence the TTVs of planet c are poorly constrained.

\begin{figure*}[ht]
		\begin{center}
			\includegraphics[width=12cm]{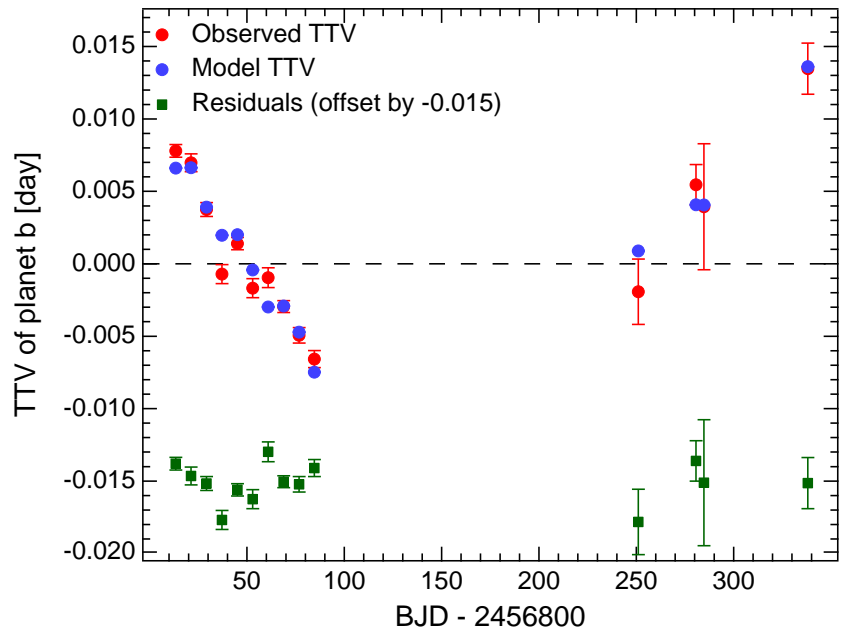} 
			\caption{Observed TTV (red) and the minimum $\chi^2$ model (blue) for the planet~b.
			Residuals are plotted with vertical offset by -0.015.
			Data for TRAPPIST and NITES are plotted with horizontal offset by -2 and +2,
			respectively, for visual purpose.
			\label{ttvplot}}
		\end{center}
	\end{figure*}

For this reason, we decide to fit only the mid-transit times of the
inner planet (14 in total) using the corresponding 4 parameters ($T_c
(0)_{b}$, $P_{b}$, $P_{c}$, and $\mu_{c}$).  This time the AMOEBA
algorithm converges.  Table~\ref{table:ttv} summarizes the optimal
parameters and Figure~\ref{ttvplot} plots the optimal model explaining
the observed TTVs of the inner planet.  We note that we find a
somewhat large reduced $\chi^2$, namely $\chi^2/\nu \sim 4.4$, where
$\nu=10$ is the number of degrees of freedom.  This may indicate that
the uncertainties in $T_{\rm c}$, particularly those based on K2 data,
have been underestimated by a factor of about 2.1.  To account for
possible systematic errors, the uncertainties given in
Table~\ref{table:ttv} have been enlarged by a factor of
$\sqrt{\chi^2/\nu}$ from the original 1$\sigma$ errors estimated by
the usual criterion $\Delta \chi^2 = 1.0$.

\begin{figure*}[ht]
		\begin{center}
			\includegraphics[width=12cm]{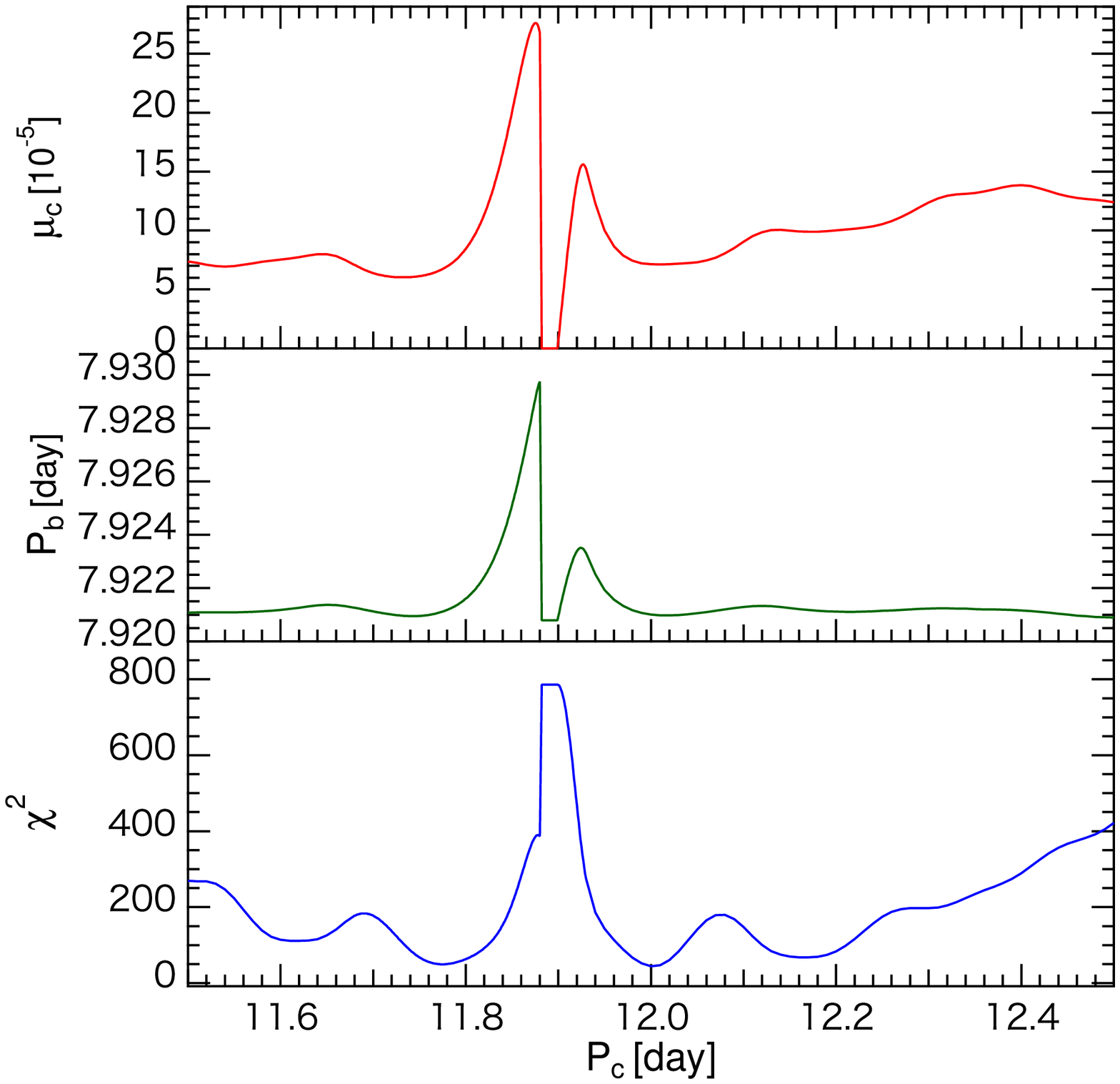} 
			\vspace{0mm}
			\caption{Variations of $\chi^2$ (bottom) and optimal $P_{\rm b}$ (middle) and $\mu_{\rm c}$ (top)
			for $P_{\rm c}$ around the minimum $\chi^2$ solution. The minimum $\chi^2$ is located at
			$P_{\rm c} = 12.0028$, while the second-minimum one is placed at $P_{\rm c} = 11.7748$
			with $\Delta \chi^2 = 5.4$. The difference of $\chi^2$ is not statistically significant.
			\label{ttvsystematic}}
			\includegraphics[width=12cm]{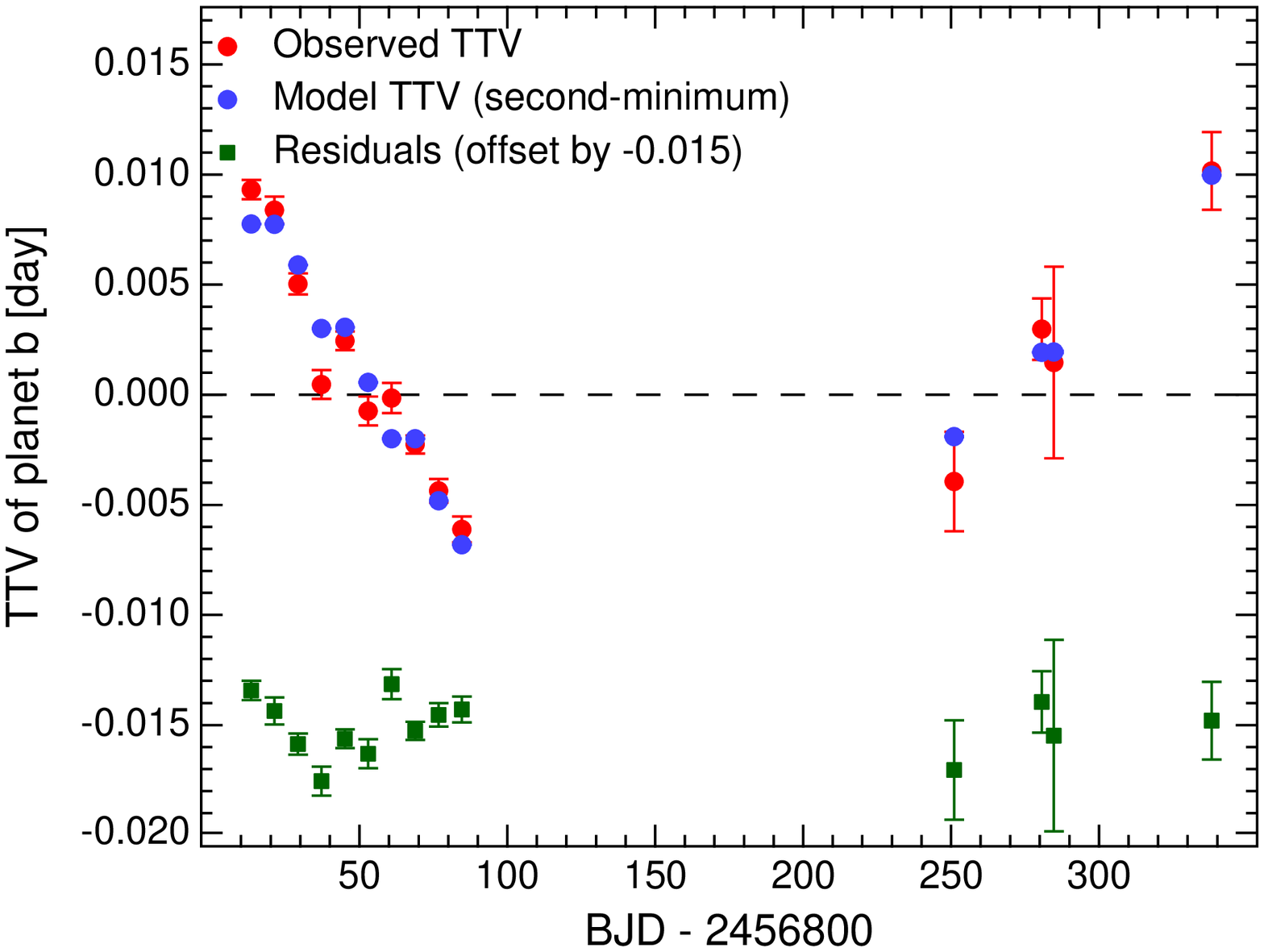} 
			\caption{Same as Figure~\ref{ttvplot}, but for the model for the second-minimum $\chi^2$.
			\label{ttvplot2}}
		\end{center}
	\end{figure*}

We then check on the existence of possible local minima of $\chi^2$ by
fixing $P_{\rm c}$ close to the optimal value and optimizing the other
parameters using the AMOEBA algorithm.  We repeat this exercise
for values of $P_{\rm c}$ ranging from 11.5 to 12.5.
Figure~\ref{ttvsystematic} shows the variation in $\chi^2$ and the optimal
parameters with changing $P_{\rm c}$.  We find that the second-minimum
$\chi^2$ is located at $P_{\rm c} \sim 11.775$ days with $\Delta
\chi^2 = 5.4$ from the minimum $\chi^2$.  Figure~\ref{ttvplot2}
presents a TTV model for the second-minimum $\chi^2$.  The behavior of
the second-minimum $\chi^2$ model is almost identical to the minimum
$\chi^2$ model in the observing period.  Thus we cannot exclude this
solution statistically at this point.  Therefore,
Table~\ref{table:ttv} also gives the parameters
and errors based on the second-minimum $\chi^2$ solution.
We also note that the third-minimum $\chi^2$
is located at $P_{\rm c} \sim 12.17$ days with $\chi^2 \sim 67.50$,
which is statistically less favorable than the above two solutions.

As seen in Figure~\ref{ttvsystematic}, the orbital period for the
inner planet $P_{\rm b}$ is robustly determined to be about 7.921
days.  The minimum and second-minimum $\chi^2$ are thus located
slightly above ($P_{\rm c}/P_{\rm b} = 1.51531 \pm 0.00117$) and below
($P_{\rm c}/P_{\rm b} = 1.48651 \pm 0.00180$) the exact 3:2
commensurability ($P_{\rm c} \sim 11.882$ days).  The orbital period
of the outer planet is not permitted to lie exactly on the
commensurability, where larger TTVs are predicted and the synodic
chopping formulae are less reliable. This is why the optimization
results in the mass of the planet c being driven to zero
when $P_{\rm c}$ is held fixed near the 3:2 commensurability. The
planet-to-star mass ratio for the outer planet $\mu_{c}$ is estimated
as $0.0000713 \pm 0.0000027$ for the minimum $\chi^2$ solution and
$0.0000672 \pm 0.0000039$ for the second-minimum.  According to the
mass of the host star presented in Table~\ref{table:HDS}, these values
correspond to a mass for planet~c of $21.4 \pm 1.9 M_{\oplus}$ and
$20.2 \pm 2.7 M_{\oplus}$ for the minimum and second-minimum
solutions, respectively.  Thus the mass of the outer planet is well
constrained to be $\sim20 M_{\oplus}$.  We note that the error of $t_{\rm
  conj}$ has less impact on the above results.  Finally, we present
coefficients of the synodic chopping formulae in Table~\ref{table:ttv}
so as to facilitate transit predictions for the inner planet in
upcoming years.

Very recently, \citet{2015arXiv151001047B} reported a photo-dynamical
modeling of TTVs and RVs of the K2-19 system.
They derived the mass of the planet c as $15.9 ^{+7.9}_{-2.8} M_{\oplus}$,
which is well consistent with our result.
We note that the derived orbital periods of both planets are also consistent
when taking account for a difference in the definitions\footnote{We define
$P_{\rm b}$ and $P_{\rm c}$ as linear coefficients
for the epoch and they are decoupled from the TTV term (see equations 8 and 9),
whereas \citet{2015arXiv151001047B} defined them as parameters in
the photo-dynamical modeling, which includes the TTV term
(see \citealt{2015arXiv151001047B}).}
of $P_{\rm b}$ and $P_{\rm c}$.

\subsection{Internal compositions of K2-19b and 19c}

\begin{figure*}[t]
		\begin{center}
			\includegraphics[width=0.65\textwidth]{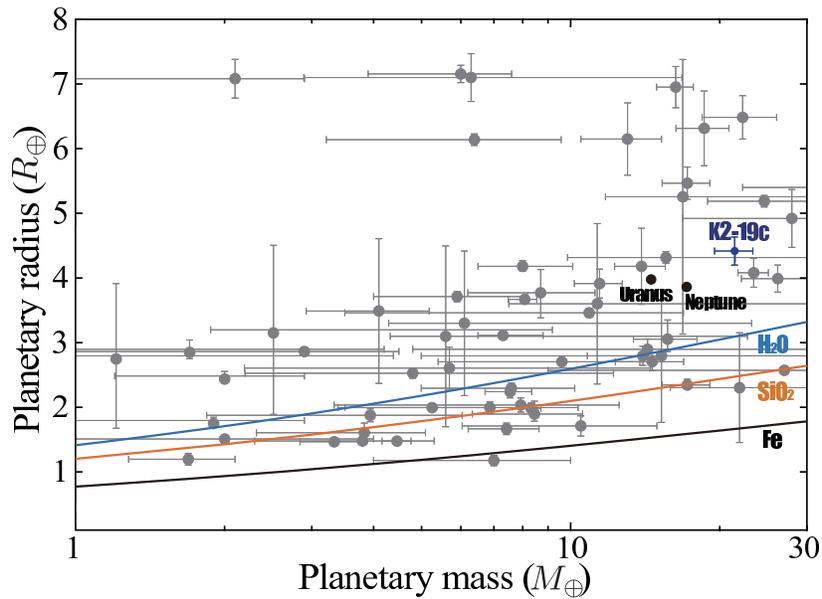} 
			\caption{
			Mass-radius relationship of transiting planets. 
			Theoretical mass-radius relations correspond to a pure water (blue), 
			a silicate (orange), and an iron (black) planet \citep{2013PASP..125..227Z}.
			K2-19c with $21.4 \pm 1.9\,M_\oplus$ (see Table~\ref{table:ttv})
			and $4.37 \pm 0.22\,R_\oplus$ (see Table~\ref{table:K2}), 
			which is larger than the radius of a pure water planet with the same mass,
			is likely to possess an atmosphere atop its core.
			Other data points shown in the figure are taken from the Exoplanet Orbit Database
			\citep{2014PASP..126..827H}.
			}
			\label{fig_MR}
		\end{center}
	\end{figure*}

K2-19c has a relatively low mean density of $1.43\,{\rm
  g\,{cm}^{-3}}$. It is therefore likely to have a gaseous atmosphere
on top of any solid component (see also Figure \ref{fig_MR}). We
explore models for the interior structure of K2-19c ($21.4\,M_\oplus$,
$4.37\,R_\oplus$ at 0.1024\,AU) based on the condition of hydrostatic
equilibrium, using equations of state (EoS) for four constituents:
SCvH EoS for H/He \citep{1995ApJS...99..713S}; the {\it ab initio} EoS
\citep{2009PhRvB..79e4107F} and the SESAME EoS 7150 \citep{1992SESAME}
for water; and the SESAME EoS 7100 and 2140 for rocky material and
iron, respectively.  We calculate a pressure-temperature profile of an
irradiated planet's atmosphere in radiative equilibrium, using the
analytical formulae of \citet{2014A&A...562A.133P} and gas opacities
derived by \citet{2008ApJS..174..504F}.  We assume that both
the incoming and outgoing radiation fields are isotropic, the bond albedo
of the planet is zero, and the incoming flux is averaged over the
dayside hemisphere.  The atmospheric mass fraction of K2-19c is
estimated to be $\sim 7\%$ for a Earth-like core model (32.5\% iron
core, 67.5\% silicate mantle) and $\sim 1.3\%$ for an icy core model
(ice : rock = 2.7 : 1), where we assumed that K2-19c has a H/He
atmosphere with the stellar metallicity of [Fe/H] = 0.07.
We find that K2-19c likely has a H/He atmosphere of $\lesssim 10$\,wt\% despite
the fact that its total mass exceeds the critical core mass for gas
accretion within the protoplanetary disk, which is thought to be
$\sim 10\,M_\oplus$ \citep[e.g.][]{1996Icar..124...62P}.  This suggests
that the atmospheric removal of K2-19c via giant impacts and/or a mass loss
driven by stellar irradiation might have occurred.

The mass of K2-19b remains undetermined.  Here, we discuss the
possible mass range of K2-19b ($7.34\,R_\oplus$ at 0.0823\,AU) from a
standpoint of the formation and evolution of close-in planets. Under
the extreme assumption that K2-19b has no gaseous atmosphere and is
purely solid, its observed radius would require its mass to be $>
100\,M_\oplus$.  The formation of such a massive solid core without
any gaseous envelope seems unlikely.  Such a massive core would be
theoretically expected to initiate the ({\it in-situ}) accretion of
the ambient disk gas \citep[e.g.][]{2012ApJ...753...66I}.  A core of
this mass has been inferred for HD~149026b
\citep{2005ApJ...633..465S}, but in that case the core is accompanied
by a substantial gaseous atmosphere.  Consequently, for K2-19b, it
seems most plausible to consider models that possess an atmosphere.
However, we note that K2-19b, orbiting at only $\sim 0.08$\,AU, may
have undergone atmospheric loss due to stellar irradiation.  As the
age of K2-19 is relatively old ($\geq 8$ Gyr), a Neptune-mass planet
could have lost a substantial amount of its atmosphere, whereas a
Jupiter-mass planet would likely have retained its original atmosphere
\citep[e.g.][]{2013ApJ...775..105O}.

With these considerations in mind, we consider two models that
K2-19b is a Neptune-like planet ($\lesssim 20$\,wt\% of a H/He atmosphere) or a gas giant. 
We first suppose that K2-19b has a
H/He atmosphere of $1\,{\rm wt}\%$, $10\,{\rm wt}\%$, and $20\,{\rm
  wt}\%$ on top of a solid core. The atmosphere is taken to have the
same metallicity as the host star, and the core is taken to have an
Earth-like composition. Under these assumptions, the mass of K2-19b is
expected to be $\sim 5.1\,M_\oplus$, $18\,M_\oplus$, and
$37\,M_\oplus$, respectively. Second, we suppose that
K2-19b has a mass of $1\,M_{\rm Jup}$.
In this case the models indicate that the
atmospheric mass fraction is $\sim 35\%$.  Accordingly, K2-19b can be
a sub-Neptune-mass planet with a tenuous atmosphere, a super-Neptune
planet, or a close-in gas giant with an extremely massive solid core.
Recently, \citet{2015arXiv151001047B} derived the mass of K2-19b as
$44\pm12$ M$_{\oplus}$, from photo-dynamical modeling.
Their result supports that K2-19b would be surrounded by
a thick H/He atmosphere of $\sim20\,{\rm wt}\%$.

\subsection{A close-in super-Neptune and Neptune pair close to the 3:2 MMR}

\begin{figure*}[t]
		\begin{center}
			\includegraphics[width=0.70\textwidth]{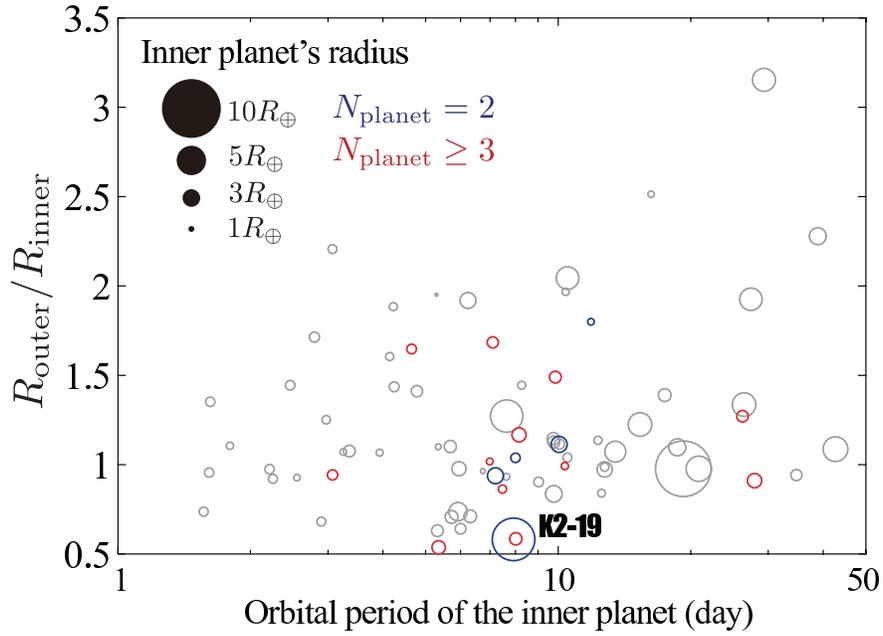} 
			\caption{Radius ratio of a neighboring pair near or in a 3:2 MMR in multiple planet systems. 
			The size of each circle is proportional to the radius of the inner planet. 
			The multiplicity of a planetary system is shown by blue (2-planet) and red ($\geq 3$\,-planet) circles.
			We also plot a pair of planets close to a 2:1 commensurability in gray.}
			\label{fig_MMR}
		\end{center}
	\end{figure*}

The K2-19 system is the first discovery of close-in Neptune-sized
planets near a 3:2 MMR.  A slow convergent migration allows low-mass
planets to be locked into a first-order commensurability
\citep[e.g.][]{2005MNRAS.363..153P}.  In fact, Figure \ref{fig_MMR}
shows that all of the other pairs of planets close to the 3:2
commensurability found by Kepler are smaller than $3\,R_\oplus$.
Also, planets with $P \lesssim 10$ days near a 2:1 MMR have radii
smaller than about $3\,R_\oplus$.  The peculiar pair close to the 3:2 MMR around
K2-19 is suggestive of inward transport of Neptune-sized planets
through a convergent migration.  We find there is a lack of close-in gas
giants among both 3:2 MMR and 2:1 MMR systems\footnote{RV surveys
  have unveiled long-period gas giants in a 3:2 MMR (HD\,45364 and
  HD\,204313) and those in a 2:1 MMR (24\,Sex, HD\,128311, HD\,37124,
  HD\,155358, HD\,73526, and HD\,82943).}.
Besides, 3:2 MMR systems seem to be less common than 2:1 MMR ones.
Dynamical instability of more closely-packed 3:2 MMR systems may be
responsible for the feature.
In addition, there is no gas giant in a 3:2 MMR so far discovered within $P = 50$ days.
For 2:1 MMR systems, gas giants such as Kepler-9b, 9c
\citep{2010Sci...330...51H}, and Kepler-30c
\citep{2012ApJ...750..114F} form the 2:1 commensurability beyond $P
\sim 10$ days.  The distinct habitat may imply that large planets with
radius of $\gtrsim 3\,R_\oplus$ most likely fall into a 2:1 MMR rather
than a 3:2 MMR due to the strength of planetary migration.

An outer planet formed in a 2:1 MMR or 3:2 MMR tends to be larger than
an inner one (see Figure \ref{fig_MMR}).
This may imply that an outer planet caught up with an inner one to form
a first-order MMR system in a protoplanetary disk.
The inner planet closer to
the central star loses its atmosphere more readily via stellar
X-ray and UV (XUV) irradiation.  The strength of the stellar XUV flux
that a pair of planets in a $p : p + 1$ MMR receives is related as
$\frac{F^{\rm out}}{F^{\rm in}} \propto
\left(\frac{p}{p+1}\right)^{4/3}$, where $p$ is the commensurability
integer and $F$ is the stellar XUV flux.  The subtle but
non-negligible flux difference may lead to the prevalence of larger
outer planets in a 2:1 MMR or a 3:2 MMR.  However, the radius of
K2-19b is about 1.68 times as large as that of the Neptune-sized
K2-19c. This means that the K2-19 system has a unique 3:2 MMR
configuration, compared to other close-in MMR systems with smaller planets.
The mass determination of K2-19b should be a crucial role of
understanding what causes their radius contrast,
and why the close-in super-Neptune and Neptune pair was trapped into
a 3:2 MMR rather than a 2:1 MMR, for example, the difference
in migration speed \citep{2013ApJ...775...34O}.

\section{Summary}

We have conducted high-dispersion spectroscopy, AO imaging, and
ground-based transit follow-up observations to characterize the K2-19
planetary system.  We have estimated parameters of the host star with
an order of magnitude higher precision than previous studies by the
Subaru/HDS spectroscopy. Through the Subaru/HiCIAO imaging, we have
excluded the presence of contaminating faint objects which can dilute
or mimic transits of the two planets. We have obtained new
ground-based transit light curves using the FLWO 1.2m telescope,
TRAPPIST 0.6m telescope, and OAO 1.88m telescope.  Combined with the
K2 transit light curves, we have modeled observed TTV for the inner
planet using the synodic chopping formulae given by
\citet{2015ApJ...802..116D}.  We have found two statistically
indistinguishable solutions for TTV, located slightly above and below
the exact 3:2 MMR.  Despite the degeneracy, both solutions
consistently estimate the orbital period of the inner planet $P_{\rm
  b} \sim 7.921$ days and the mass of the outer planet $M_{\rm c} \sim
20 M_{\oplus}$.

To break the degeneracy of the TTV solutions, further ground-based
transit follow-up observations as well as precise radial-velocity
measurements are strongly desired.  We particularly
emphasize the importance of transit observations for the outer planet,
which will allow us to characterize the TTV of the outer planet in
detail and thereby enable us to estimate the mass of the inner planet
precisely.  Given the nearly-equatorial location of K2-19 in the sky
(RA: 11:39:50.477, Dec: +00:36:12.87, \citealt{2013AJ....145...44Z}),
the K2-19 system is observable from both the northern and southern
hemispheres. This will facilitate intensive follow-up observations,
which would be quite fruitful in the upcoming seasons.  A precise
estimate of the mass of the inner planet will allow us to estimate the
amount of a H/He atmosphere, for which we have presented theoretical
predictions based on different internal compositions.  The nature and
the amount of the H/He atmosphere for the inner planet can be further
characterized by transmission spectroscopy.  Since the inner planet is
a rare case of a super-Neptune, such a characterization of the
atmosphere would be also interesting as future work.

\acknowledgments N.N. acknowledges supports by the NAOJ Fellowship,
Inoue Science Research Award, and Grant-in-Aid for Scientific Research
(A) (JSPS KAKENHI Grant Number 25247026).  T.H. is supported by Japan
Society for Promotion of Science (JSPS) Fellowship for Research
(No. 25-3183). A.F. acknowledges supports by the Astrobiology Center Project of
National Institutes of Natural Sciences (NINS) (Grant Number AB271009).
Y.H. is supported by Grant-in-Aid for Scientific
Research on Innovative Areas (No. 26103711) from MEXT.  This work was
performed, in part, under contract with the Jet Propulsion Laboratory
(JPL) funded by NASA through the Sagan Fellowship Program executed by
the NASA Exoplanet Science Institute.  Work by J.N.W. was supported by
the NASA Origins program (grant NNX11AG85G).  M.G. and E.J. are
Research Associates at the Belgian Fund for Scientific Research (Fonds
National de la Recherche Scientifique, F.R.S-FNRS).  L.D. received the
support of the F.R.I.A. fund of the FNRS.  TRAPPIST is a project
funded by the F.R.S-FNRS under grant FRFC 2.5.594.09.F, with the
participation of the Swiss National Science Foundation (SNF).
M.T. is supported by Grant-in-Aid for Scientific Research (No.15H02063).
This research has made use of the Exoplanet Orbit Database
and the Exoplanet Data Explorer at exoplanets.org.
We also thank Masahiro Ogihara for helpful comments on the dynamics of MMR systems.

{\it Facilities:} \facility{Subaru}, \facility{OAO}, \facility{FLWO}, \facility{TRAPPIST}, \facility{NITES}.

\clearpage
%%%%%%%%%%%%%%%%%%%%%%%%%%%%%%%%%%%%%%%%%%%%%%%%%%%%%%%%%%%%%%%%%%%%%%
%%\bibliography{refs}

%%%%%%%%%%%%%%%%%%%%%%%%%%%%%%%%%%%%%%%%%%%%%%%%%%%%%%%%%%%%%%%%%%%%%%

\clearpage
%%%%%%%
%%%%%%%

%%%%%%%
%%%%%%%

%%%%%%%
%%%%%%%

%%%%%%%
%%%%%%%

%%%%%%%
%%%%%%%
				
%%%%%%%
%%%%%%%

%%%%%%%
%%%%%%%

%%%%%%%
%%%%%%%

\clearpage
%\vspace{-3cm}
%%%%%%%%%%%%%%%%%%%%%%%%
\begin{table}
\begin{center}
\caption{Stellar parameters derived from high dispersion spectroscopy.\label{table:HDS}}
\begin{tabular}{lcc}
\tableline\tableline
parameter & value & error$^a$ \\ \tableline
$T_{\rm eff}$ [K] & 5345 & 17 \\
$\log g$ & 4.394 & 0.050 \\
$[{\rm Fe}/{\rm H}]$ & 0.07 & 0.03 \\
$R_{\rm s}$ [$R_{\odot}$] & 0.914 & 0.027 \\
$M_{\rm s}$ [$M_{\odot}$] & 0.902 & 0.011 \\
$V \sin I_{\rm s}$ [km~s$^{-1}$] & 0.85 & $^{+0.95}_{-0.85}$ \\
Age [Gyr] & $\geq 8$ & -- \\
\tableline\tableline
\end{tabular}
\tablenotetext{a}{Presented errors are derived from statistical measurement errors in equivalent widths of iron lines.
Additional systematic error (say, 40 K in $T_{\rm eff}$) may exist.
Discussions on such systematic errors are presented in \citet{2014ApJ...783....9H}.}
\end{center}
\end{table}
%%%%%%%%%%%%%%%%%%%%%%%%

%%%%%%%%%%%%%%%%%%%%%%%%
\begin{table}
\begin{center}
\caption{Best fitting parameters and errors for K2 transit light curves.
Derived planetary radii are also presented.\label{table:K2}}
\begin{tabular}{lcccccc}
\tableline\tableline
% & \multicolumn{6}{c}{common}  \\
parameter & \multicolumn{2}{c}{} & value & error  & \multicolumn{2}{c}{} \\ \tableline
$a/R_{\rm s,b}$ & \multicolumn{2}{c}{} & 19.35  & $^{+0.56}_{-1.45}$  & \multicolumn{2}{c}{} \\
$a/R_{\rm s,c}$ & \multicolumn{2}{c}{} & 24.09  & $^{+1.72}_{-3.01}$  & \multicolumn{2}{c}{} \\
$b_{\rm b}$ & \multicolumn{2}{c}{} & 0.233  & $^{+0.213}_{-0.163}$  & \multicolumn{2}{c}{} \\
$b_{\rm c}$ & \multicolumn{2}{c}{} & 0.367  & $^{+0.260}_{-0.246}$  & \multicolumn{2}{c}{} \\
$R_{\rm p}/R_{\rm s,b}$ & \multicolumn{2}{c}{} & 0.0737  & $^{+0.0016}_{-0.0011}$  & \multicolumn{2}{c}{} \\
$R_{\rm p}/R_{\rm s,c}$ & \multicolumn{2}{c}{} & 0.0439  & $^{+0.0018}_{-0.0009}$  & \multicolumn{2}{c}{} \\
$T_{\rm 14,b}$ [day] & \multicolumn{2}{c}{} &  0.1362 & $^{+0.0020}_{-0.0014}$  & \multicolumn{2}{c}{} \\
$T_{\rm 14,c}$ [day] & \multicolumn{2}{c}{} &  0.1536 & $^{+0.0042}_{-0.0027}$  & \multicolumn{2}{c}{} \\
$u_1 + u_2$ & \multicolumn{2}{c}{} &  0.69 & $^{+0.18}_{-0.14}$  & \multicolumn{2}{c}{} \\
$u_1 - u_2$ & \multicolumn{2}{c}{} & 0.06  & $^{+0.35}_{-0.38}$  & \multicolumn{2}{c}{} \\
\tableline
$R_{\rm p,b}$ [$R_{\oplus}$] & \multicolumn{2}{c}{} & 7.34  & $\pm 0.27$  & \multicolumn{2}{c}{} \\
$R_{\rm p,c}$ [$R_{\oplus}$] & \multicolumn{2}{c}{} & 4.37  & $\pm 0.22$  & \multicolumn{2}{c}{} \\
\tableline\tableline
\end{tabular}
\end{center}
\end{table}
%%%%%%%%%%%%%%%%%%%%%%%%

%%%%%%%%%%%%%%%%%%%%%%%%
\begin{table*}
\begin{center}
\caption{A list of mid-transit times ($T_{\rm c}$) for planet b and c from K2 transit light curves.\label{table:K2Tc}}
\begin{tabular}{cccc|cccc}
\tableline\tableline
\multicolumn{4}{c}{planet b}  & \multicolumn{4}{c}{planet c} \\ \tableline
epoch & $T_{\rm c}$ & error & $\beta$ & epoch & $T_{\rm c}$ & error & $\beta$ \\ \tableline
0 & 2456813.38403 & 0.00044 &  1.00 &
0 & 2456817.27227 & 0.00144 &  1.00 \\
1 & 2456821.30421 & 0.00061 &  1.00 &
1 & 2456829.18406 & 0.00170 &  1.12 \\
2 & 2456829.22197 & 0.00048 &  1.12 &
2 & 2456841.09346 & 0.00198 &  1.30 \\
3 & 2456837.13851 & 0.00066 &  1.00 &
3 & 2456853.00224 & 0.00214 &  1.03 \\
4 & 2456845.06161 & 0.00042 &  1.00 &
4 & 2456864.90713 & 0.00127 &  1.00 \\
5 & 2456852.97953 & 0.00066 &  1.03 &
5 & 2456876.81473 & 0.00135 &  1.04 \\
6 & 2456860.90124 & 0.00068 &  1.00 &
6 & 2456888.71247 & 0.00191 &  1.00 \\
7 & 2456868.82024 & 0.00041 &  1.00 &
 & & & \\
8 & 2456876.73925 & 0.00054 &  1.04 &
 & & & \\
9 & 2456884.65861 & 0.00059 &  1.00 &
 & & & \\
\tableline\tableline
\end{tabular}
\end{center}
\end{table*}
%%%%%%%%%%%%%%%%%%%%%%%%

%%%%%%%%%%%%%%%%%%%%%%%%
\begin{center}
\begin{table*}[t]
\caption{Best fitting parameters and uncertainties for ground-based transit light curves.\label{table:ground}}
\begin{tabular}{lcccccc}
\tableline\tableline
& \multicolumn{2}{c}{FLWO} & \multicolumn{2}{c}{TRAPPIST} & \multicolumn{2}{c}{NITES} \\
parameter & value & error & value & error & value & error \\ \tableline
epoch & \multicolumn{2}{c}{30} & \multicolumn{2}{c}{34} & \multicolumn{2}{c}{34} \\
$T_{\rm c}$
& 2457051.00413 & $^{+0.00218} _{-0.00225}$ 
& 2457082.69550 & $^{+0.00140} _{-0.00116}$
& 2457082.69398 & $^{+0.00342} _{-0.00435}$ \\
$R_{\rm p}/R_{\rm s}$ & 0.0633 & $^{+0.0042} _{-0.0045}$ & 0.0682 & $^{+0.0043} _{-0.0046}$ & 0.0645 & $^{+0.0082} _{-0.0109}$ \\
$i$ [deg] & 90.00 & $^{+0.80} _{-0.83}$ & 90.00 & $\pm 0.46$ & 90.00 & $\pm 0.72$ \\
$a/R_{\rm s}$ & 19.22 & $^{+0.44} _{-0.56}$ & 19.58 & $^{+0.30} _{-0.34}$ & 19.35 & $^{+0.43} _{-1.12}$ \\
$k_0$ & 0.9899 & $\pm 0.0025$ & 1.0012 & $\pm 0.0017$ & 0.9982 & $^{+0.0011} _{-0.0013}$\\
$k_t$ & -0.0077 & $\pm 0.0024$ & \multicolumn{2}{c}{--} & -0.0043 & $^{+0.0031} _{-0.0035}$ \\
$k_z$ & 0.00199 & $^{+0.00075} _{-0.00079}$ & -0.00020 & $^{+0.00036} _{-0.00035}$ & \multicolumn{2}{c}{--}\\
$k_{\rm flip}$ & \multicolumn{2}{c}{--} & -0.00030 & $^{+0.00033} _{-0.00031}$ & \multicolumn{2}{c}{--}\\
$\beta$ & \multicolumn{2}{c}{1.258} & \multicolumn{2}{c}{1.000} & \multicolumn{2}{c}{1.000}\\ \hline \hline
& \multicolumn{2}{c}{MuSCAT(g$^\prime _2$)} & \multicolumn{2}{c}{MuSCAT(r$^\prime _2$)} & \multicolumn{2}{c}{MuSCAT(z$_{\rm s,2}$)}\\
parameter & value & error & value & error & value & error \\ \tableline
epoch & \multicolumn{2}{c}{--} & \multicolumn{2}{c}{41} & \multicolumn{2}{c}{--} \\
$T_{\rm c}$ & \multicolumn{2}{c}{--} & 2457138.15047 & $^{+0.00145} _{-0.00176}$ & \multicolumn{2}{c}{--} \\
$R_{\rm p}/R_{\rm s}$ & 0.0753 & $^{+0.0067} _{-0.0070}$ & 0.0789 & $^{+0.0043} _{-0.0041}$ & 0.0730 & $^{+0.0058} _{-0.0061}$ \\
$i$ [deg] & \multicolumn{2}{c}{--} & 88.94 & $^{+0.73} _{-0.95}$ & \multicolumn{2}{c}{--}\\
$a/R{\rm s}$ & \multicolumn{2}{c}{--} & 18.80 & $^{+0.94} _{-2.10}$ & \multicolumn{2}{c}{--}\\
$k_0$ & 0.9943 & $^{+0.0046} _{-0.0047}$ & 0.9945 & $\pm 0.0028$ & 0.9913 & $\pm 0.0040$\\
$k_t$ & -0.0243 & $^{+0.0046} _{-0.0047}$ & -0.0107 & $\pm 0.0027$ & -0.0095 & $\pm 0.0041$\\
$k_z$ & 0.0014 & $\pm 0.0015$ & 0.00188 & $\pm 0.00089$ & 0.0030 & $\pm 0.0013$\\
% $k_{\rm flip}$ & \multicolumn{2}{c}{--} & \multicolumn{2}{c}{--} & \multicolumn{2}{c}{--}\\
$\beta$ & \multicolumn{2}{c}{1.059} & \multicolumn{2}{c}{1.000} & \multicolumn{2}{c}{1.069}\\
\tableline\tableline
\end{tabular}
\end{table*}
\end{center}
%%%%%%%%%%%%%%%%%%%%%%%%

%%%%%%%%%%%%%%%%%%%%%%%%
\begin{table*}
\begin{center}
\caption{Planetary parameters derived by the TTV analysis.
Coefficients for the synodic chopping formulae are also presented.\label{table:ttv}}
\begin{tabular}{lcccc}
\tableline\tableline
& \multicolumn{2}{c}{minimum $\chi^2$} & \multicolumn{2}{c}{second minimum $\chi^2$}\\
parameter & value & error$^a$ & value & error$^a$ \\ \tableline
$P_{\rm b}$ [day] & 7.920994 & 0.000071 &  7.921122 & 0.000176 \\
$P_{\rm c}$ [day] & 12.0028 & 0.0092  &  11.7748 & 0.0142 \\
$\mu_{\rm c}$ & 0.0000713 & 0.0000057  &  0.0000672 & 0.0000087 \\
$T_c (0)_{b}$ & 2456813.37624 & 0.00050  &  2456813.37466 & 0.00109 \\
$\chi^2 / \nu$ & $43.98 / 10$ & -- & 49.38 / 10  & -- \\
\tableline \tableline
$M_{\rm c}$ & 21.4 & 1.9 & 20.2 & 2.7 \\
$P_{\rm c}/P_{\rm b}$ & 1.51531 & 0.00117 &  1.48651 & 0.00180 \\
$f_b^{(1)}(\alpha)$ & 11.91 & -- & 13.51  & -- \\
$f_b^{(2)}(\alpha)$ & 20.09 & -- & 21.19  & -- \\
$f_b^{(3)}(\alpha)$ & -139.71 & -- & 178.84  & -- \\
$f_b^{(4)}(\alpha)$ & -4.10 & -- & -5.56  & -- \\
$f_b^{(5)}(\alpha)$ & -1.22 & -- & -1.57  & -- \\
$f_b^{(6)}(\alpha)$ & -0.50 & -- & -0.64  & -- \\
$P_{\rm syn}$ & 23.29234 & -- & 24.20293 & -- \\
$\alpha$ & 0.75799 & -- & 0.76776 & -- \\
\tableline \tableline
\end{tabular}
\tablenotetext{a}{Presented errors are inflated by $\sqrt{\chi^2 / \nu}$ to account for possible systematic errors.}
\end{center}
\end{table*}
%%%%%%%%%%%%%%%%%%%%%%%%

\end{document}